\documentclass[iop, twocolumn, appendixfloats]{emulateapj}

\usepackage{apjfonts}
\usepackage{natbib}
\usepackage{amsfonts}
\usepackage{amsmath}
\usepackage{amssymb}
\usepackage[usenames]{color}
\usepackage{verbatim}
\usepackage{longtable}
\usepackage{array}
\usepackage{epsfig}
\usepackage{scrextend}
\usepackage{mathrsfs}
\usepackage{rotating}
\newcommand{\code}[1]{\texttt{#1}}

\newcommand{\mesa}{\code{MESA}}
\newcommand{\MESA}{\mesa}

\newcommand{\mesastar}{\mesa~\code{star}}
\newcommand{\MESAstar}{\mesastar}

\newcommand{\intone}{\code{interp\_1d}}

\newcommand{\unitstyle}[1]{\ensuremath{\mathrm{#1}}}

\newcommand{\Msun}{\ensuremath{\unitstyle{M}_\odot}}

\input{macros/vectors}
\newcommand{\Teff}{\ensuremath{T_{\!\mathrm{eff}}}}	

\newcommand{\Tc}{\ensuremath{T_{\mathrm{\!c}}}} 
\newcommand{\rhoc}{\ensuremath{\rho_{\mathrm{c}}}} 
\newcommand{\Minit}{\ensuremath{\mathrm{M_{init}}}}

\usepackage[colorlinks=true,urlcolor=blue]{hyperref}

\begin{document}
\title{\MESA\ Isochrones and Stellar Tracks (MIST) 0:\\ Methods for the construction of stellar isochrones}

\shorttitle{Isochrone construction}
\shortauthors{A.\ Dotter}

\author{Aaron Dotter} 
\affil{Research School of Astronomy and Astrophysics\\ Australian National University\\
  Canberra, ACT\\ Australia}
\email{aaron.dotter@gmail.com} 

\begin{abstract}
I describe a method to transform a set of stellar evolution tracks onto a uniform basis and then interpolate within 
that basis to construct stellar isochrones. The method accommodate a broad range of stellar types, from substellar 
objects to high-mass stars, and phases of evolution, from the pre-main sequence to the white dwarf cooling sequence. 
I discuss situations in which stellar physics leads to departures from the otherwise monotonic relation between initial
stellar mass and lifetime and how these may be dealt with in isochrone construction. I close with convergence tests
and recommendations for the number of points in the uniform basis and the mass between tracks in the original grid
required in order to achieve a certain level accuracy in the resulting isochrones. The programs that implement these 
methods are free and open-source; they may be obtained from the project webpage.\footnote{\url{https://github.com/dotbot2000/iso}}
\end{abstract}


\section{Introduction}\label{s:intro}

To borrow a line from the poet John Godfrey Saxe, isochrones, like sausages, cease to inspire respect in proportion as we know how they are made.
Nevertheless, the intent of this paper is to explain one method of isochrone construction and show the results of the codes that implement the method.

A stellar evolution code produces output at fixed points in time---timesteps---that allow one to follow the evolution of a model 
star over some portion of its lifetime. Timesteps are chosen to meet various numerical tolerances and may vary by orders of 
magnitude over the span of the evolutionary sequence. The resulting data per timestep constitutes a stellar evolution track, which is
a fundamental tool in the study of stellar evolution. The primary input parameters of a stellar evolution track are its 
initial mass (\Minit) and chemical composition but there are a host of other details relating to the physics and numerics assumed in any model.

\begin{figure*}
\includegraphics[width=\textwidth]{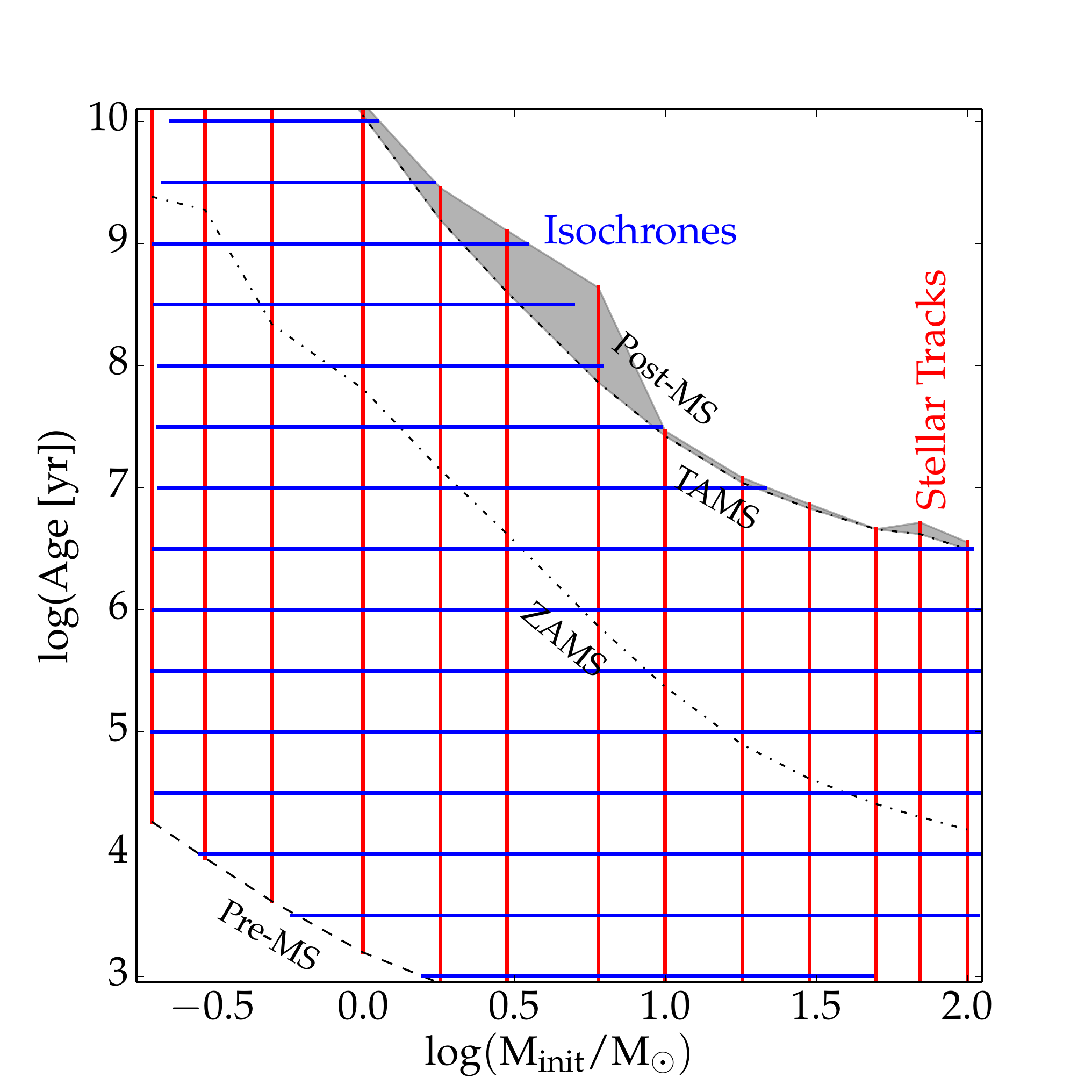}
\caption{Schematic view of stellar evolution tracks and isochrones in the \Minit-age plane. The diagonal lines and corresponding labels mark
the positions of evolutionary phases up to and including the main sequence. The gray shaded region shows all post-main sequence evolution.}
\label{fig:mass-age}
\end{figure*}

An isochrone is another, complimentary tool that is useful when the properties of a stellar population---rather than a single star---are of interest. 
An isochrone is derived from a set of stellar evolution tracks spanning a range of \Minit\ but with the same initial chemical
composition. The underlying assumption in the creation of an isochrone is that all stars are formed simultaneously from a 
homogeneous gas cloud. (Whether or not these assumptions are ever \emph{actually} met is beyond the scope of this paper.) 
The goal of isochrone construction is to change the independent variable from \Minit\ in the set of stellar evolution tracks to time in the isochrones, as illustrated in Figure \ref{fig:mass-age}.

A representation of stellar evolution tracks and isochrones in the same plane is shown in Figure \ref{fig:mass-age}. 
The isochrones (horizontal lines) have been generated from the tracks (vertical lines) using the methods described in this
paper. It should be noted, however, that the isochrones were generated from a denser grid of tracks than is
shown here for clarity. The thin lines running diagonally through Figure \ref{fig:mass-age} identify three different phases
of stellar evolution (see $\S$\ref{ss:secondary}): the pre-main sequence (`Pre-MS'), the zero age main sequence (`ZAMS'),
and the terminal age main sequence (`TAMS'). The shaded region above the TAMS line indicates the post-MS evolution.

The process of isochrone construction is trivial if the stellar models all have lifetimes far exceeding the desired isochrone age. 
In this case, all that is required is a simple interpolation within each stellar evolution track to the desired age. 
This is evident in Figure \ref{fig:mass-age} where multiple tracks intersect with a given isochrone before and during the main sequence (MS).
An example of such are the Lyon models of very-low-mass stars \citep{Baraffe1997,Baraffe1998}. 
The problem becomes much harder when we wish to capture the late phases of stellar evolution. Referring again to Figure \ref{fig:mass-age}, 
the entirety of the post-MS may lie between two stellar tracks in the model grid, see the gray shaded region in the figure. In this case, we
encounter a situation in which neighboring stellar evolution tracks in the model grid are in completely different phases of evolution 
and a sophisticated approach is required to construct isochrones that faithfully reproduce the morphology of the tracks. Many of the isochrone 
libraries in use today, such as BaSTI \citep{Pietrinferni2004}, Dartmouth \citep{Dotter2008},
PARSEC \citep{Bressan2012}, Pisa \citep{Tognelli2011}, Victoria-Regina \citep{VandenBerg2006}, and Yale-Yonsei \citep{Yi2001}, are based 
on the more-sophisticated approach, one implementation of which is described in the following sections. 
The literature describing isochrone construction is sparse considering their widespread use in astrophysics; the most notable example is 
the work of P.\ Bergbusch and D.\ VandenBerg \citep{Bergbusch1992,Bergbusch2001,VandenBerg2006,VandenBerg2012}.

\begin{figure*}
\includegraphics[width=\textwidth]{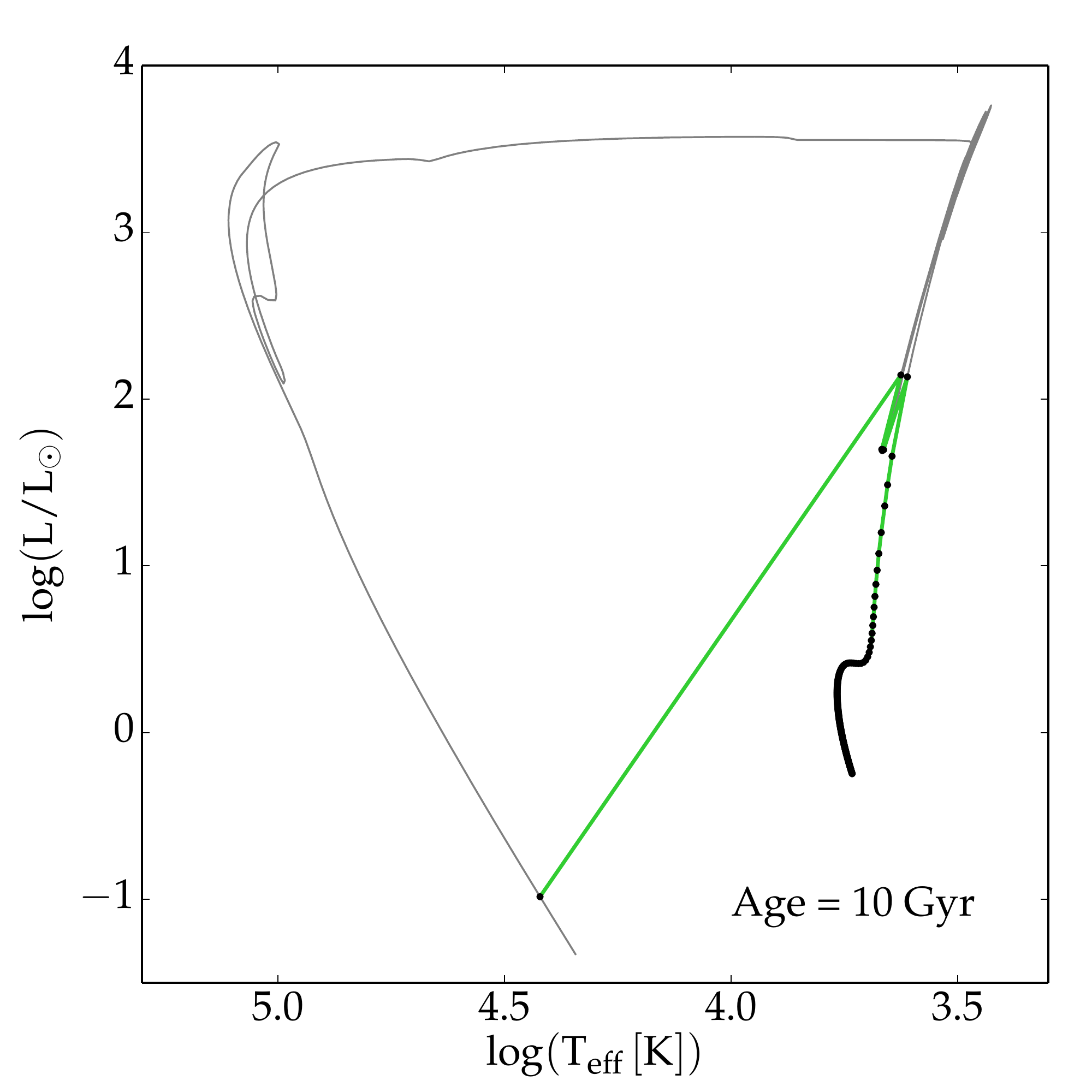}
\caption{The filled circles show the points at which a grid of stellar evolution tracks reach 10 Gyr. The difference in \Minit\ between each 
successive point is 0.001 \Msun. The green line shows the simple approach to isochrone construction in which a given age is located in each 
track and then these points are connected by line segments. The gray line shows an isochrone made using the sophisticated approach described 
in this paper.\label{fig:naive}}
\end{figure*}

A common question: Isn't it possible to make isochrones from a sufficiently-finely-sampled grid of stellar evolution tracks?  Then isochrone construction 
requires a trivial interpolation to `connect the dots' for a given age. The short answer is that such an approach is both inefficient and inelegant. 
Inefficient because the number of tracks required is orders of magnitude greater than is typically seen ($10^4$-$10^5$ compared to hundreds).
Inelegant because there is a great deal of similarity between two stellar evolution tracks with comparable \Minit; a sophisticated approach to isochrone 
construction will exploit the similarity.
An example is provided in Figure \ref{fig:naive} where isochrones were constructed using this simple approach as well as the more-sophisticated 
approach described later in this paper from a grid of stellar evolution tracks with mass sampling $\Delta$M=0.001 \Msun\ between 0.85 and 2.15 \Msun\ 
(1301 tracks in total).\footnote{This mass sampling is much finer than is typically found in stellar evolution libraries, where 0.05 \Msun\ sampling is 
typical for this range of stellar masses.} The stellar evolution tracks in this grid were evolved to the white dwarf cooling sequence (WDCS) and 
the isochrones show the full range of evolution in the tracks.

The agreement between the simple and the sophisticated in Figure \ref{fig:naive} is decent along the main sequence but begins to break down on 
the red giant branch (RGB). The simple approach skips from a point on the RGB to the core He-burning phase, bypassing the upper RGB, and then
from core He-burning to near the bottom of the WDCS. Put another away, the difference in \Minit\ between the early AGB and the WDCS is only 0.001 \Msun!
Yet this interval covers a great deal of distance in the H-R diagram and includes the brightest stars in the isochrone. 

\begin{figure*}
\includegraphics[width=\textwidth]{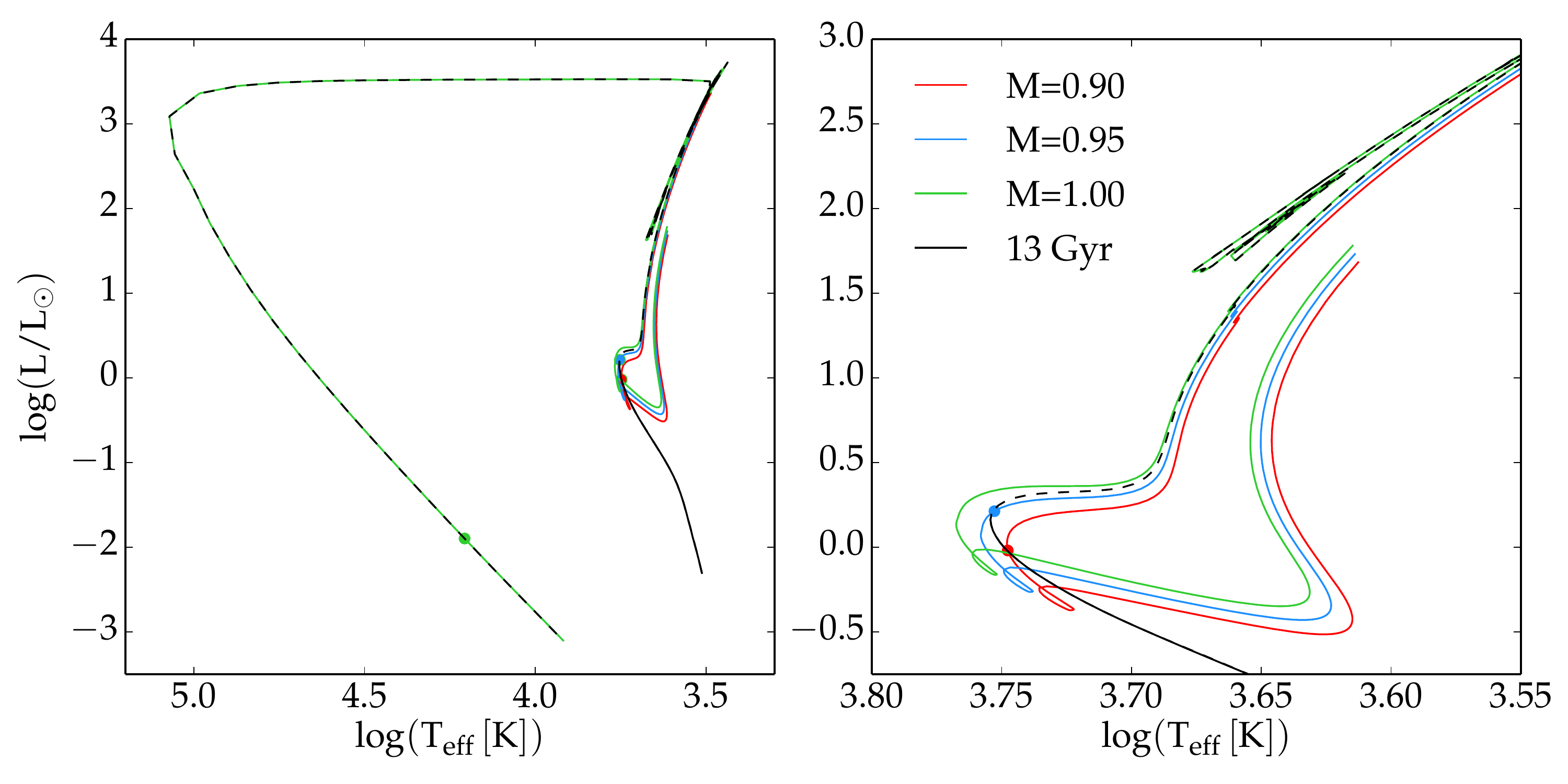}
\caption{A 13 Gyr isochrone and three stellar evolution tracks with $\Minit = 0.90, 0.95, 1.00~\Msun$.
The left panel shows the entire span of a 13 Gyr isochrone (black) and the $1~\Msun$ track (green). The 0.90 (red) and $0.95~\Msun$ (blue)
tracks are only plotted from the pre-main sequence to the end of the RGB phase for clarity. The right panel focuses on the main sequence, subgiant 
branch, and red giant branch. The isochrone is plotted as a solid line for all points that satisfy $M_{init} < 0.95~\Msun$ and as a dashed line 
elsewhere. Finally, the point at which each track has an age of 13 Gyr is shown as a filled circle; for the $1~\Msun$ track this point is only 
visible in the left panel.}
\label{fig:iso-tracks}
\end{figure*}

As a more in-depth example, consider that we wish to construct a 13 Gyr isochrone from a set of stellar evolution tracks with a mass 
interval of $0.05~\Msun$ as visualized in Figure \ref{fig:iso-tracks}. The $1.0~\Msun$ track has a MS lifetime of 10 Gyr and a total 
lifetime of 14 Gyr. The $0.95~\Msun$ track has a MS lifetime of 12.3 Gyr and a total lifetime of 16.5 Gyr. The $0.90~\Msun$ track has 
a MS lifetime of 15.1 Gyr and a total lifetime of 20 Gyr. At 13 Gyr the $0.9~\Msun$ model is still on the MS, the $0.95~\Msun$ 
model is a subgiant, and the $1~\Msun$ model is on the WDCS. The simple approach of locating the 
point at which each track has an age of 13 Gyr and `connecting the dots' will result in an isochrone that jumps directly from the
MS to the WDCS, skipping over all the intervening evolutionary phases. If instead we use the approach outlined in the following sections,
then we obtain the isochrone shown in Figure \ref{fig:iso-tracks}, which faithfully reproduces all of the evolutionary phases. The isochrone shown 
in Figure \ref{fig:iso-tracks} closely follows the $1 \Msun$ track from the red giant branch through to the end because the later phases 
are relatively short-lived.

The stellar evolution tracks used throughout this paper are taken from \MESA\ Isochrones and Stellar Tracks (MIST; Choi et al., ApJS, submitted) 
and were computed with the stellar evolution code \MESAstar, part of the \MESA\ code 
library \citep{Paxton2011,Paxton2013,Paxton2015}.\footnote{\url{http://mesa.sourceforge.net}}
The programs that implement the methods described in the following sections are written in Fortran and use \MESA\ modules, primarily for 
interpolation. Although originally designed to ingest \MESAstar\ history files, it is certainly possible to incorporate tracks 
from other stellar evolution codes. It is only a matter of loading the tracks into the data structures used in the codes.

\begin{figure*}
\includegraphics[width=\textwidth]{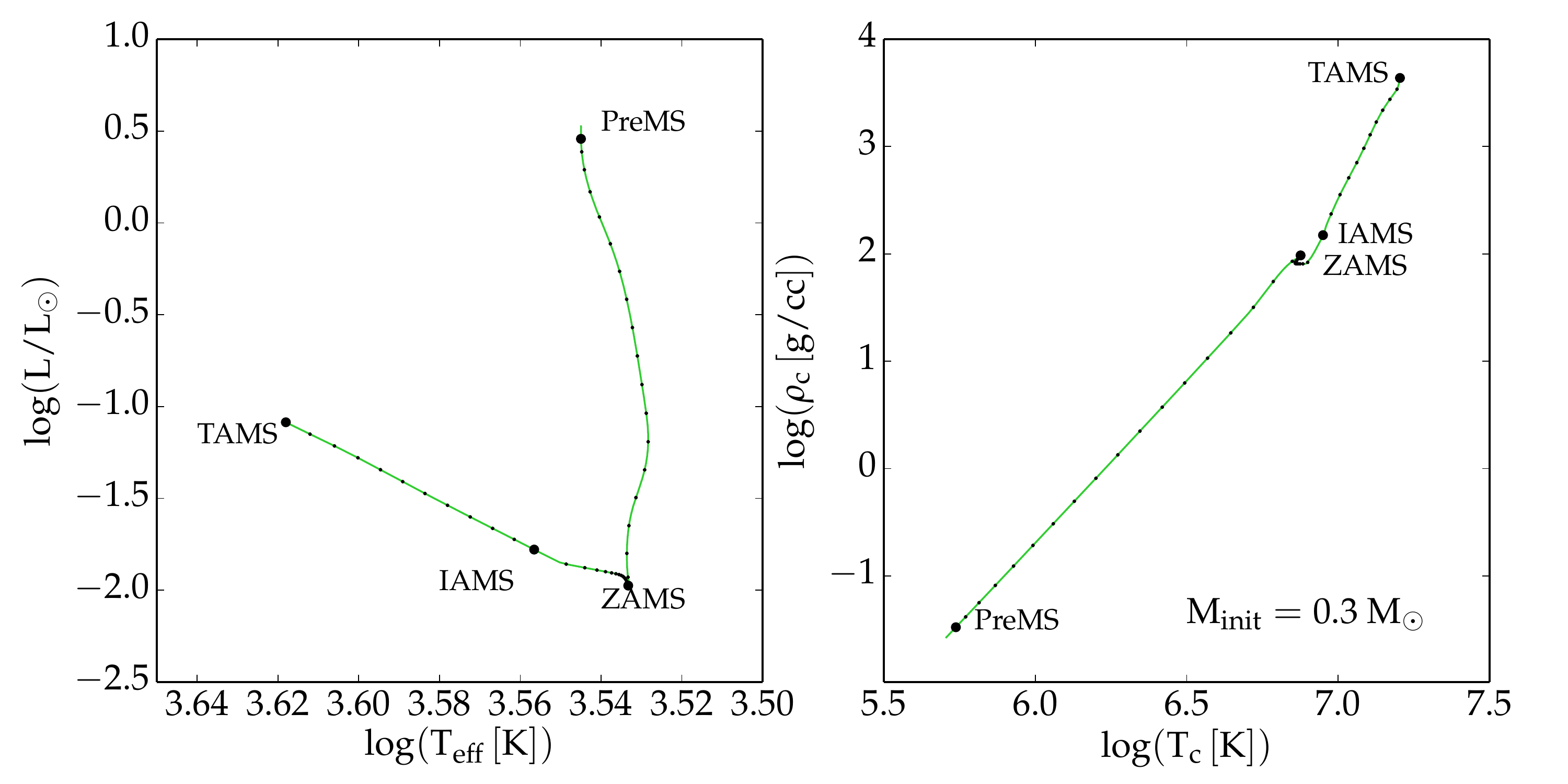}
\caption{A $0.3~\Msun$ stellar evolution track in the H-R diagram (right) and the \Tc-\rhoc\ diagram (left).
The original track is shown as the solid line. The primary EEPs described in 
$\S$\ref{ss:primary} are labeled and marked by large dots; secondary EEPs described in $\S$\ref{ss:secondary} are shown as small dots along the track.\label{fig:03M}}
\end{figure*}

\begin{figure*}
\includegraphics[width=\textwidth]{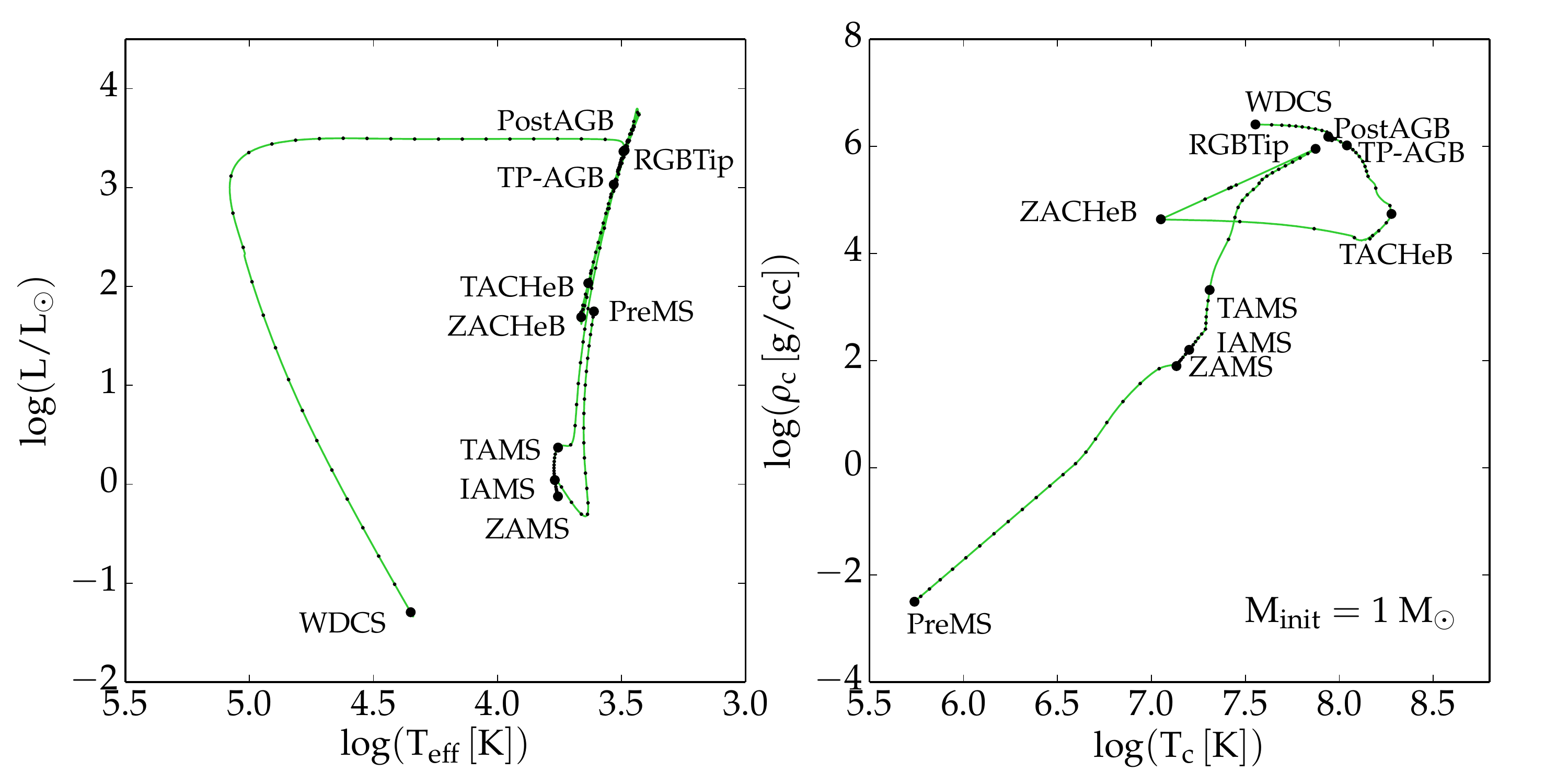}
\caption{Equivalent to Figure \ref{fig:03M} for a $1~\Msun$ stellar evolution track.\label{fig:1M}}
\end{figure*}

\begin{figure*}
\includegraphics[width=\textwidth]{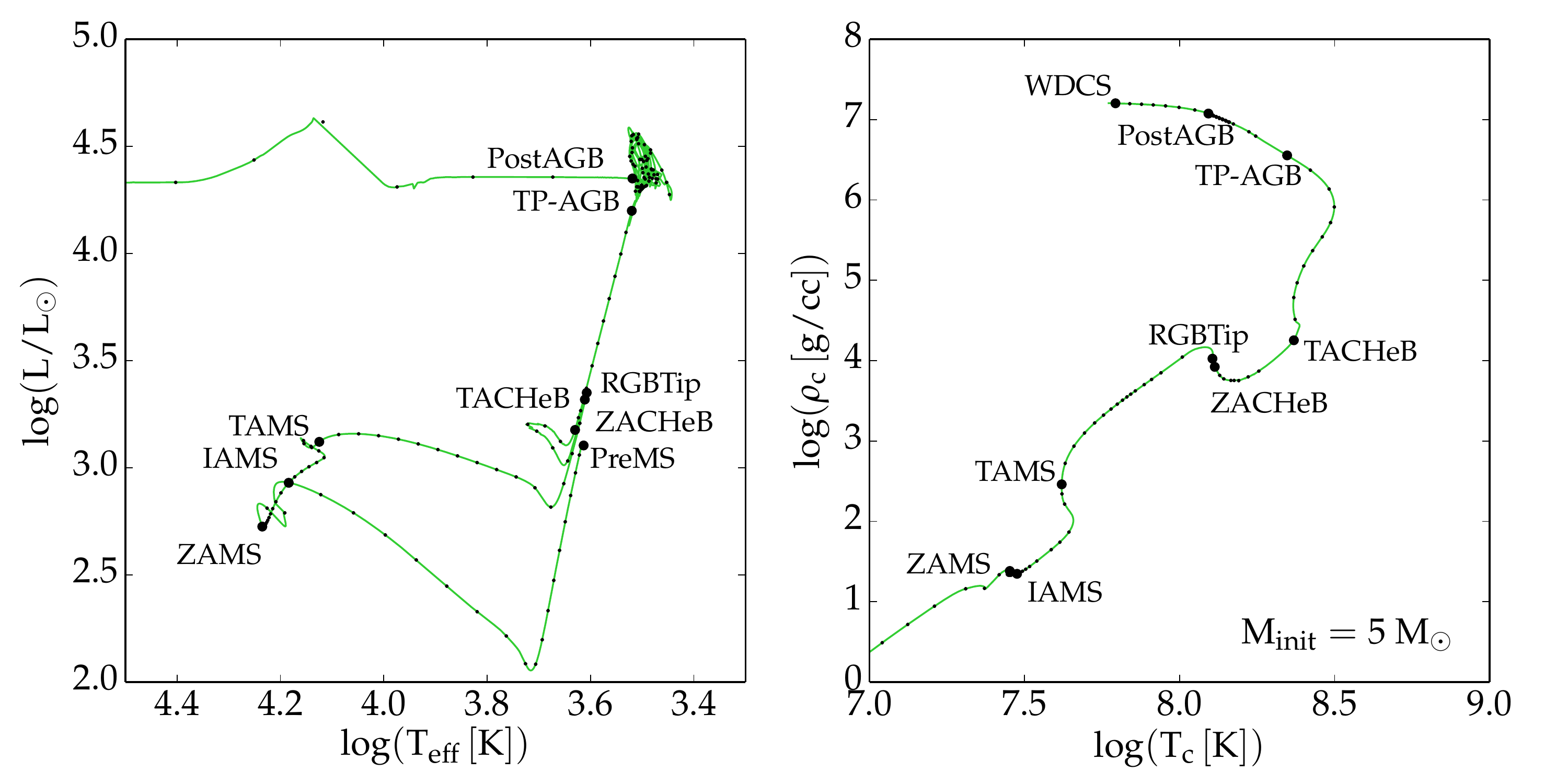}
\caption{Equivalent to Figure \ref{fig:1M} for a $5~\Msun$ model. Not all primary EEPs are shown in both panels.\label{fig:5M}}
\end{figure*}

\begin{figure*}
\includegraphics[width=\textwidth]{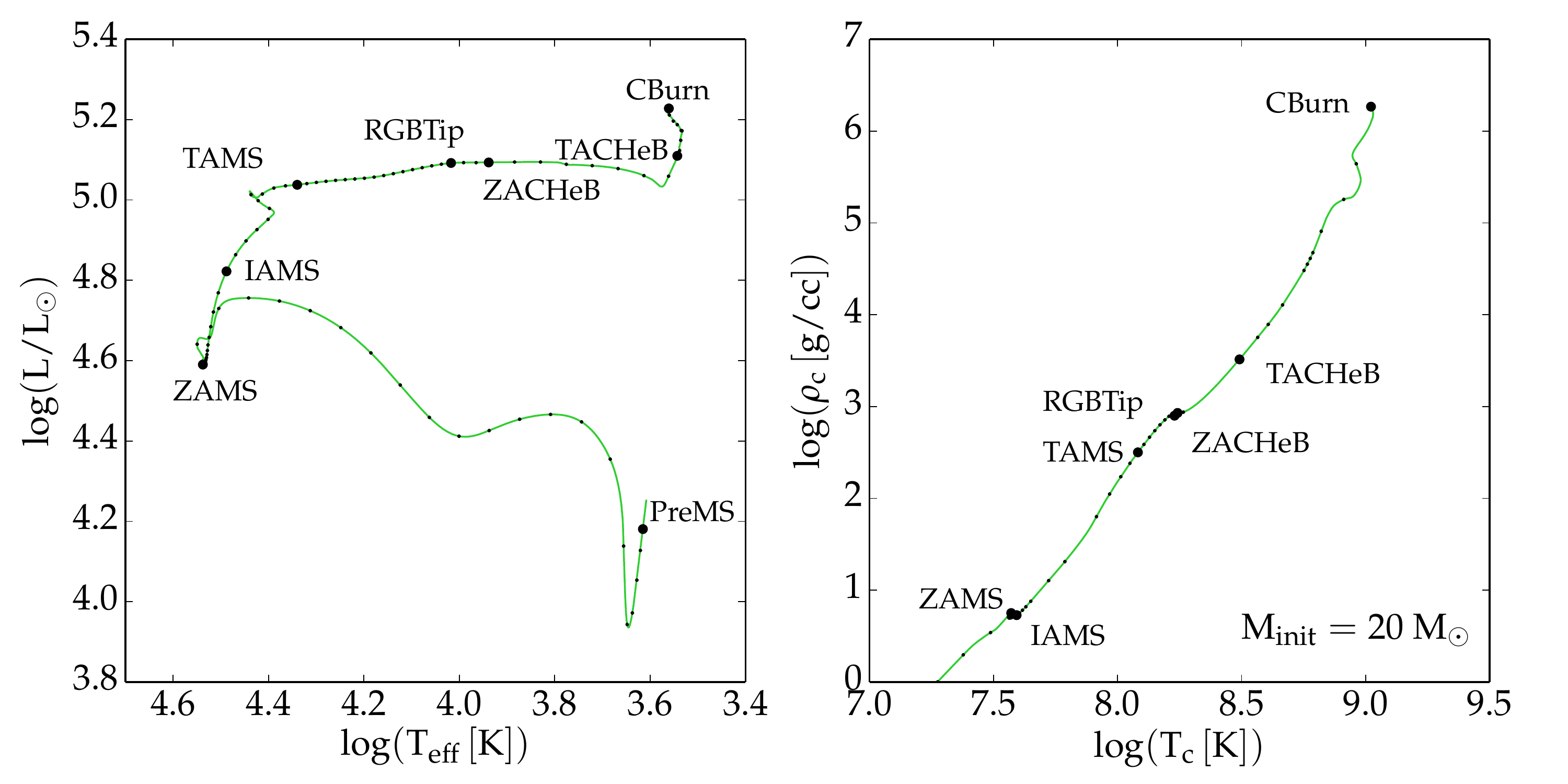}
\caption{Equivalent to Figure \ref{fig:1M} for a $20~\Msun$ model. Not all primary EEPs are shown in both panels. 
Note that in the \Tc-\rhoc\ plane two pairs of primary EEPs are nearly coincident: ZAMS and IAMS; RGBTip and ZACHeB.  The RGBTip primary EEP does not have the same significance in such high-mass stars but is maintained for consistency, as noted in the text.\label{fig:20M}}
\end{figure*}

Interpolation plays a key role in isochrone construction.  It is therefore worthwhile 
to describe the key features of the interpolation method used throughout the following sections. A cubic 
interpolation scheme provides a good balance between smoothness, including continuous first derivatives, 
and a reasonably small number of neighboring points required to construct the interpolating function. 
One important consideration is that the interpolation method preserve monotonicity throughout 
the interpolation interval. This is particularly important in isochrone construction because the standard procedure
hinges on a monotonic relation between \Minit\ and age.
The piecewise-monotonic cubic interpolation method of \citet{steffen_1990_aa}, as implemented in 
the \MESA\ \intone\ module, meets both of these criteria and is used by default in these codes.  
All references to interpolation throughout the remainder of this paper refer to the \citet{steffen_1990_aa} method.

Units in this paper are generally cgs, except as explicitly noted, and ages are always given in years.
The following sections present the theory and relevant details of how the codes are implemented;
a practical guide to using the programs is distributed along with the codes themselves.

\section{Identifying equivalent evolutionary phases}\label{s:eep}
The complexities of stellar structure and evolution imply that stellar evolution tracks spanning a
range of \Minit\ are likely to have (perhaps vastly) different lifetimes and numbers of timesteps. 
The model stars may experience entirely different evolutionary phases. This situation is far from ideal 
if the goal is to interpolate amongst the tracks to construct isochrones.

The task of interpolating amongst a set of stellar evolution tracks can be greatly simplified by 
transforming the original tracks, as output by the stellar evolution code, onto a uniform basis. 
The uniform basis is designed in such a way that each phase of stellar evolution
is represented by a fixed number of points and that the $n$th point in one track has a comparable interpretation in another track.
This is accomplished by introducing the concept of \emph{equivalent evolutionary phases} (EEPs),
a series of points that can be identified in all stellar evolution tracks.\footnote{A description of isochrone construction using central H mass fraction as a time analog by \citet{Simpson1970} is, to my knowledge, the earliest discussion in the literature of something like the EEP formalism. 
}
EEPs serve as the uniform basis to describe the evolution of all stars and are divided into two categories.
\emph{Primary} EEPs identify a relatively small number of physically-motivated phases in the tracks.
\emph{Secondary} EEPs provide a uniform spacing between the primary EEPs in each track.

\begin{figure*}
\includegraphics[width=0.8\textwidth]{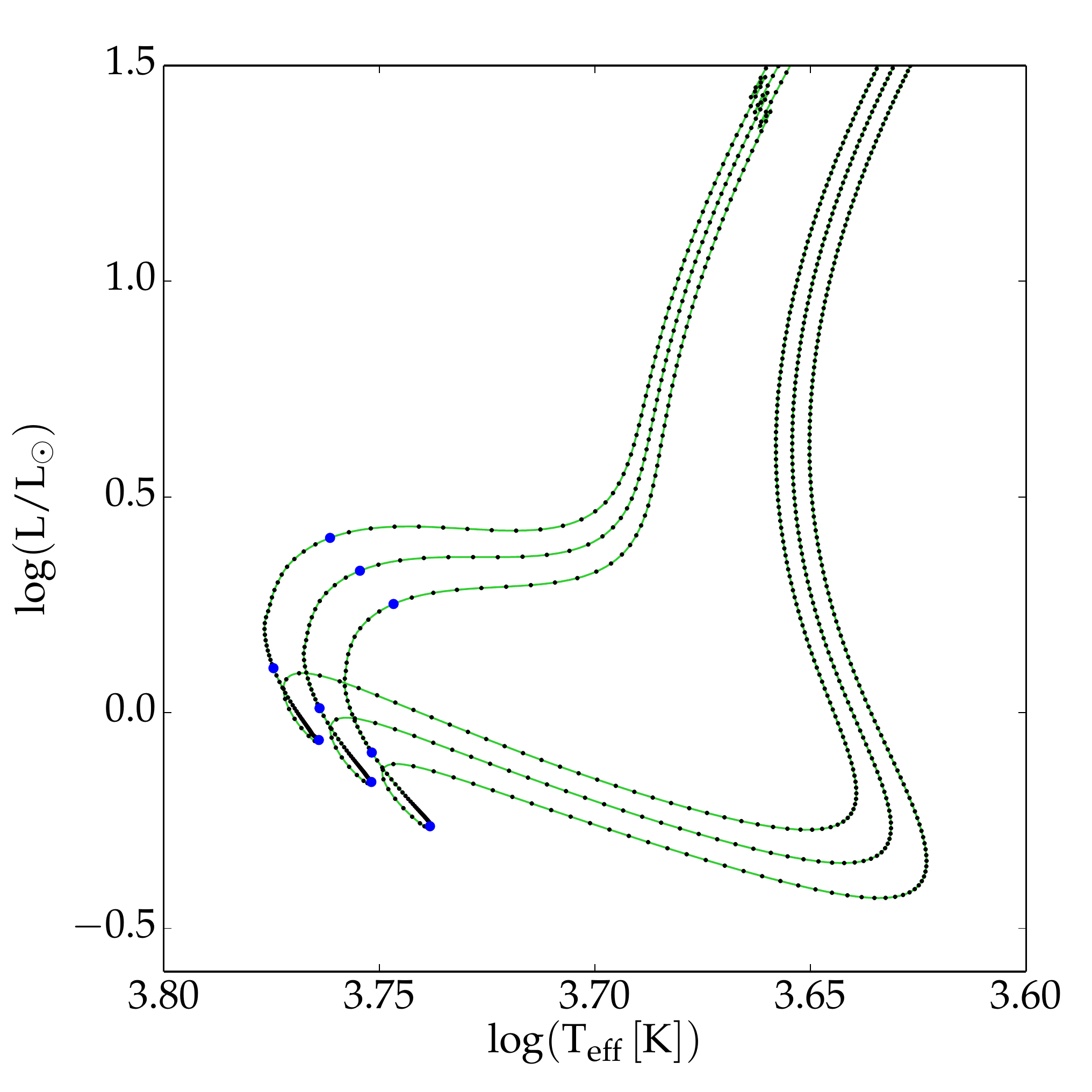}
\caption{This figure shows how the primary and secondary EEPs are distributed along three stellar evolution tracks
with \Minit = 0.95, 1, and $1.05~\Msun$.  The original tracks are plotted as solid lines; the secondary EEPs are shown as
small dots and the primary EEPs are shown as large dots. This should give some indication that the EEP-based tracks 
constitute a uniform basis.\label{fig:eep_track}}
\end{figure*}

\subsection{Primary EEPs}\label{ss:primary}
\emph{Primary} EEPs are explicitly defined for each track and have the same physical 
interpretation across different tracks. However, since the goal of the codes described here is
to be applicable to all stars, there is one branch point (see EEP 8 in the list below). The
distinction between low- and intermediate-mass stars on one hand and high-mass stars on the other
is determined by whether or not the central temperature of the model at the end of its evolutionary 
track is less than the central temperature after the end of core He-burning. The $\Tc$ criterion is an 
indication of whether the models star will evolve to a cooling WD (in which case $\Tc$ is lower at the end)
or proceed on to later stages of nuclear burning (in which case $\Tc$ is higher at the end). Furthermore, 
substellar objects, for which substantial nuclear burning does not occur, are treated differently. 

This section lists all of the primary EEPs and explains how each is defined.
In what follows $X$ and $Y$ refer to the mass fractions of H and He, respectively, and subscript $c$ 
denotes the property at the center of the star. For example, $X_c$ is the central H mass fraction.
Examples of 0.3, 1, 5, and $20~\Msun$ stellar tracks with primary EEPs identified are shown in 
Figures \ref{fig:03M}, \ref{fig:1M}, \ref{fig:5M}, and \ref{fig:20M}, respectively. Note that the number of EEPs
shown in Figures \ref{fig:03M}-\ref{fig:20M} has been reduced by a factor of 10 from the recommended numbers
so that the individual EEPs may be clearly seen.

\begin{itemize}
\item[1.] The pre-MS EEP (PreMS) is chosen to identify the point at which the central 
temperature ($\Tc$) rises above a certain value (lower than necessary for sustained nuclear reactions). 
By default this is set at $\log(\Tc)=5.0$ but may be set to a different value. 
It is recommended that the pre-MS point be chosen as early as possible.
If the first point in the stellar evolution track is already above this threshold, 
we simply use the first point. It is implicitly assumed that even substellar objects, nominally
$M \la 0.08 \Msun$, will meet the PreMS $\Tc$ condition.

\item[2.] The zero-age main sequence (ZAMS) EEP is taken as the first point after the H-burning 
luminosity exceeds 99.9\% of the total luminosity \emph{and} before the central H mass fraction 
has fallen by below its initial value by 0.0015. In the substellar case where neither of these 
criteria is met, the ZAMS point is taken as the maximum in $\Tc$ along the evolutionary track.

\item[3.-4.] Towards the end of core H-burning 2 primary EEPs are defined at $X_c=0.3$ (intermediate 
age main sequence: IAMS) and $X_c=10^{-12}$ (terminal age main sequence: TAMS). Two points are used 
to isolate the portion of the MS that may or may not be influenced by the presence of a 
convective core. The practical choice of placing two primary EEPs near the end of the MS 
greatly simplifies the treatment of the `convective hook' feature in the H-R diagram. In the substellar
case, for which no central H is consumed, the final point in the track is chosen as the TAMS provided
that the age at that point is greater than some minimum (e.g., 20 Gyr).

\item[5.] The RGB tip (RGBTip) EEP identifies the point at which the stellar luminosity reaches a maximum---or 
the stellar $\Teff$ reaches a minimum---after core H burning is complete but before core He burning has progressed 
significantly. This EEP has a recognizable location on the H-R diagram of low- and intermediate mass stars, hence 
the name, but the point defined here can also be located in high-mass stellar tracks that do not go through a `red giant' phase.
This is achieved by taking the point at which the luminosity reaches a maximum or the $\Teff$ reaches a minimum, 
whichever comes first, before the center He mass fraction ($Y_c$) is significantly reduced by He burning: $Y_c \ge Y_{c,TAMS}-0.01$.

\item[6.] The zero age core He burning (ZACHeB) EEP denotes the onset of sustained core He burning.
The point is identified as the $\Tc$ minimum that occurs after the onset of 
He-burning (RGBTip) while $Y_c > Y_{c,RGBTip}-0.03$. This temperature minimum 
is readily identifiable in lower-mass stars ($\Minit < 2 \Msun$) because the \Tc-\rhoc\ evolution 
in this interval has a particular shape due to the off-center ignition of He burning under degenerate
conditions \citep{Paxton2011}. The same feature is less obvious in higher-mass stars with 
non-degenerate cores but still identifiable.

\item[7.] One primary EEP is identified at the end of core He burning (terminal age core He burning: 
TACHeB) corresponding $Y_c=10^{-4}$.

\item[8a.] The EEP marking the onset of the thermally-pulsing AGB (TP-AGB) is identified as the point 
after core He burning ($Y_c < 10^{-6}$) when the difference in mass between the H-burning and He-burning 
shell is less than $0.1\Msun$. This is the same criterion used in \MESAstar\ to identify the onset of thermal pulsations. 

\item[8b.] For stellar models that are massive enough to bypass the TP-AGB and proceed to later
phases of core burning, the final EEP is set at the end of core C burning (CBurn), when the central C
mass fraction falls below $10^{-4}$. This marks the end of the primary EEPs for massive stars. The 
remaining primary EEPs are only applicable to low- and intermediate-mass stars.

\item[9.] A post-AGB (PostAGB) EEP is identified only in stellar models that will go on to form a WD. 
It is meant to locate the point at which the TP-AGB phase has ended and the star
has begun to cross the H-R diagram at nearly constant luminosity. The PostAGB EEP is defined as 
the point at which the H-rich stellar envelope falls below 20\% of the current stellar mass.

\item[10.] The WDCS EEP, which follows the Post-AGB, is based on the central value of the Coulomb 
coupling parameter $\Gamma$, with a default upper limit of 100. WDCS is only considered for models 
that have a PostAGB EEP.
\end{itemize}

The code processes each primary EEP in order using the location of the previous EEP as starting point 
and searching for the conditions needed to identify the next EEP through to the end of the track.
The process ends either when the full list of primary EEPs have been identified or at the first 
failure to identify a primary EEP. The list presented above is the complete list of primary EEPs
in the code.  However, it is not necessary to use all of them.  For example, one could skip the PreMS
primary EEP for a grid of stellar evolution tracks that begin from the ZAMS.

\subsection{Secondary EEPs}\label{ss:secondary}
\emph{Secondary} EEPs serve the purpose of faithfully capturing the morphology of each segment
of the evolutionary track that lies between two adjacent primary EEPs. After the primary EEPs have 
been identified, each segment is populated with a number of equally-spaced secondary EEPs. 
In order to construct a uniform basis of EEP-based tracks for interpolation, the number of secondary 
EEPs between a given pair of primary EEPs is held constant over the set of stellar evolution tracks.
Figures \ref{fig:03M} through \ref{fig:20M} show both primary (as larger dots) and secondary
EEPs (as smaller dots). Furthermore, Figure \ref{fig:eep_track} shows 3 neighboring tracks from a
model grid with the primary and secondary EEPs shown; in this case it is easy to see how the EEPs in
one track correspond to those in each of the others and, thus, form a suitable basis for interpolation.

Once the primary EEPs have been identified the stellar evolution track is ready to be converted
from its original form to the new form consisting only of EEPs. The new, EEP-based track consists
of all the primary EEPs that were identified in the original track plus a larger number of secondary
EEPs. Using a metric function calculated along the original 
track ($\S$\ref{ss:metric}), each interval between 2 primary EEPs is divided into a fixed number of 
equally-spaced secondary EEPs; the information to be included in the EEP-based track is interpolated 
from the original track onto to the secondary EEPs. Processing stellar evolution tracks in this way results 
in a reduction in the size of the evolutionary tracks by a factor of $\sim$10 without significant loss 
of information for a suitable number of secondary EEPs (see $\S$\ref{ss:convergence}).

\subsection{The metric function}\label{ss:metric}
For the secondary EEPs to be `equally-spaced' we must define a metric along the evolutionary track. 
The only (mathematical) constraint on the metric is that it must be positive definite. Traditionally 
the metric has been defined as a Euclidean distance along the stellar evolution track in the 
Hertzsprung-Russell (H-R) diagram. However, it can be useful to include an age term with appropriate weight
\citep[see the discussion in $\S$4 of][]{VandenBerg2012}. The terms are weighted (stretched) such that the logarithmic 
ranges spanned by a stellar evolution track in luminosity and \Teff\ contribute to the metric
distance in roughly equal amounts. Additional terms, with arbitrary weights, may be added as the user wishes. 
The metric distance $D$ along the track is defined such that $D_0=0$ and the distance between any two points 
$i$ and $i+1$ in the original evolutionary track are given by
\begin{equation}\label{eq:metric}
D_{i+1} = D_i + \sqrt{ \sum\limits_{j=1}^N w_j (x_{j,i+1} - x_{j,i})^2 }.
\end{equation}
Then the total distance along the track is $\sum_i D_i$ but, in practice, the quantity of interest
is the distance \emph{between} two adjacent primary EEPs.
The $w_j$ in Equation \ref{eq:metric} are arbitrary weight factors applied to the individual terms.
The $x_j$ are columns of data from the stellar evolution track. It is convenient, but not strictly necessary, 
to use logarithmic quantities for the $x_j$ because many of the quantities used span an order of magnitude or 
more. In principle, any quantity from the stellar evolution track may be used in Equation \ref{eq:metric}.
Experiments with central temperature and density indicate that both can be used effectively in the 
metric funcion. It should be noted that the singular purpose of the metric is to place secondary EEPs 
along the evolutionary track between two primary EEPs; the metric is not used explicitly in any later steps.

\subsection{Summary}\label{ss:summary}

Section \ref{s:eep} describes a two-step process to assign EEPs and create uniform, EEP-based stellar evolution tracks
that can later be used for interpolation of other tracks and isochrones. The first step is to cycle through the original 
stellar evolution track and identifies the points that are coincident with the primary EEPs. The second step is to cycle 
through the original track again and locate the desired number of secondary EEPs between each pair of primary EEPs. 
It is important to note that although primary and secondary EEPs are distinct in the way that they are \emph{identified} in 
the original stellar evolution tracks there is no difference in the way they are \emph{used} for interpolation purposes.
The total number of points in the EEP-based track will equal the sum of the number of primary EEPs and the number of 
secondary EEPs between each pair of primaries.

\section{Constructing new tracks and isochrones}\label{s:interp}
Once a set of stellar evolution tracks has been processed onto a uniform grid of EEPs, it is 
straightforward to interpolate new stellar evolution tracks for \Minit\ values not included 
in the original grid and to generate isochrones; extrapolation is not permitted.

\subsection{Creating a new stellar evolution track}\label{ss:track}
The process of generating a new stellar evolution track is as simple as identifying a subset
of 2 or more (4 for cubic interpolation) tracks in the original grid whose \Minit\ values 
envelop that of the new track. Once a set of tracks with appropriate \Minit-values have been indentified,
a new track can be created by looping over all EEPs, interpolating amongst those tracks at fixed EEP number
using \Minit\ as the independent variable.
The resulting track has the same number of EEPs as those that went into constructing it.

\begin{figure*}
\includegraphics[width=\textwidth]{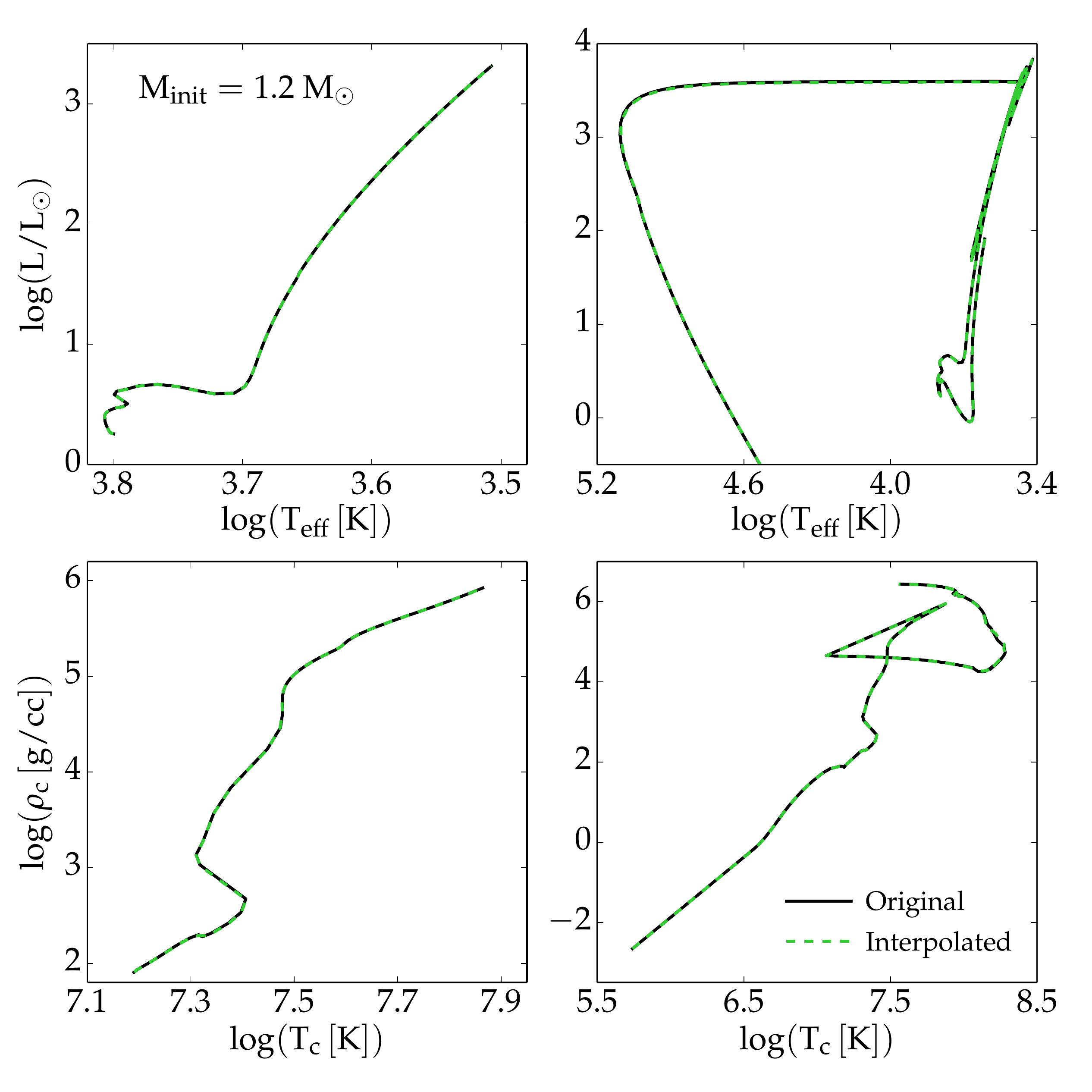}
\caption{Comparison of a $1.20~\Msun$ stellar evolution track created by interpolation with one 
computed directly by \MESAstar.}
\label{fig:interp-mass}
\end{figure*}

Consider the example shown in Figure \ref{fig:interp-mass}, where a new stellar evolution track
with $\Minit=1.2~\Msun$, created by cubic interpolation from 4 tracks with \Minit = 1.10, 1.15, 
1.25, and 1.30 \Msun, is compared with the actual \Minit = 1.2 \Msun\ track from the existing set 
of models. Figure \ref{fig:interp-mass}
shows both the H-R diagram (top row) and the central temperature-density diagram (bottom row).
Panels on the left show the full evolutionary tracks, from pre-MS to a cooling white dwarf (WD),
while on the right are shown the evolution from ZAMS to RGBTip. 

\subsection{Creating isochrones}\label{ss:isochrone}

The process of generating an isochrone is more involved than that of generating a new track
but it is, nevertheless, straightforward on a uniform grid of EEPs. For a given age, an isochrone
is constructed by looping over the complete set of EEPs, identifying which are valid for that
age, and then performing the necessary interpolations to arrive at the stellar parameters for that EEP.
In this context, an EEP is `valid' if more than one evolutionary track exists at that age for 
that EEP. For example, an ancient isochrone (age $>$ 10 Gyr) will have no valid EEPs on the pre-MS
while a very young isochrone (age $<$ 5 Myr) will have no TP-AGB or WDCS.

The construction of an isochrone for a given age is performed as a loop over all EEPs:
\begin{enumerate}
\item Make a list of the stellar evolution tracks that include each EEP at that age. If none exist, then cycle to the next EEP.
\item For that EEP and list of tracks, an ordered \Minit-age relation is constructed. 
\item Using the input age, obtain the \Minit-value appropriate for that age and EEP by interpolation in the \Minit-age relation (see Figure \ref{fig:nonmono}).
\item With \Minit\ in hand, all stellar parameters for that EEP are obtained by another round of interpolations using \Minit\ as the independent variable.
\end{enumerate}
The whole process is then repeated for other ages.

Instances, primarily during the TP-AGB phase, can occur when interpolation and even finite numerical precision can lead to non-monotonic 
\Minit\ values along a given isochrone. Such instances are numerical rather than physical and, thus, distinct from those 
discussed in $\S$\ref{ss:complications} and \ref{ss:double}. In this case it is possible to enforce monotonicity via the 
`pool adjacent violators' (PAV) algorithm.\footnote{\url{http://stat.wikia.com/wiki/Isotonic\_regression}}
The PAV algorithm is an iterative procedure that enforces monotonicity by searching through an array and replacing a non-monotonic value 
with a weighted average of neighboring points.


\subsection{Mitigating non-monotonic behavior in the \Minit-age relation}\label{ss:complications}

The standard assumption employed in isochrone construction is that, for a given EEP, the \Minit-age relation will
be monotonic. Thus for each age and EEP a unique \Minit\ value is obtained. Each EEP in a given isochrone should, therefore, have a 
higher \Minit\ than the one before it. However, stellar physics is not always so conformant and there are instances where, over a 
small interval in \Minit, the \Minit-age relation is non-monotonic. 
The mathematical formalism described above does not work in such cases. In fact, it is possible to miss the signature of such
transitions if the grid of stellar evolution tracks is sufficiently coarse. The finer the grid, the more pronounced the effect
\citep{Girardi2013}.

\begin{figure*}
\includegraphics[width=\textwidth]{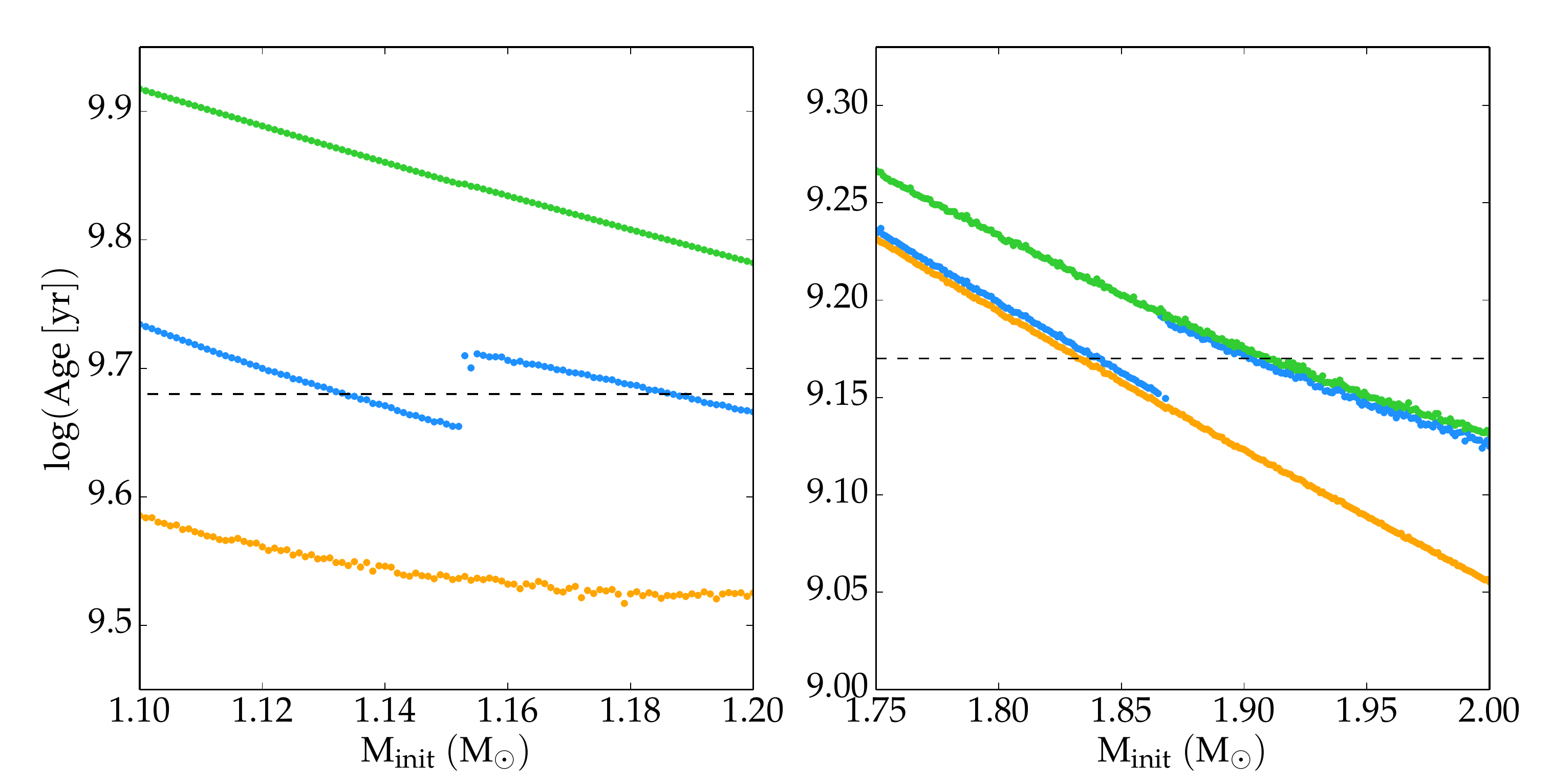}
\caption{Transitions in stellar physics lead to nonmonotonic \Minit-age relations in (at least) two cases. This figure demonstrates two cases where non-monotonicity occurs.  Both panels show the \Minit-age relations of three successive EEPs, denoted by different colors; the first and last EEPs (orange and green, respectively, in both panels) exhibit normal, monotonic behavior. The intermediate EEP (blue) exhibits non-monotonic behavior.
\emph{Left:} The 
transition from radiative to convective core on the MS. \emph{Right:} The transition between degenerate and non-degenerate 
onset of core He-burning \citep{Girardi2013}. Both panels show a sequence of EEPs during which the nonmonotonic behavior appears.
The dashed horizontal lines in each panel indicate isochrone ages for which multiple stars would appear in the same EEP (if allowed to).
\label{fig:nonmono}}
\end{figure*}

Even with careful choice of primary EEPs ($\S$\ref{ss:primary}) 
and distance metric ($\S$\ref{ss:metric}), changes in stellar physics can lead to non-monotonic behavior in the \Minit-age relation.
There are two well-known examples of this: the first is the appearance of the convective core during core H-burning in stars
more massive than $1\Msun$. The second is the transition between the onset of core He-burning under degenerate and non-degenerate conditions
around 1.8-$2\Msun$ \citep{Girardi2013}.\footnote{A third, weaker case is the transition from fully-convective stars to stars with
radiative cores, which takes place around $0.35\Msun$. Indeed, there may be other cases but the two referred to in the text are
the most pronounced examples.} In both cases, the change in stellar physics leads to a marked difference in stellar 
lifetime that temporarily reverses the trend of decreasing lifetime with increasing \Minit.

Figure \ref{fig:nonmono} gives an example of such non-monotonic behavior towards the end of the MS due to the appearance of a convective
core (left panel) and the case presented by \citet[][right panel]{Girardi2013}. Shown in the left panel are the \Minit-age relations for MS EEPs. 
Shown in the right panel are \Minit-age relations for core He-burning EEPs. In either case, the intermediate EEP is noticeably non-monotonic. 
An isochrone for a particular range of ages (suggested by the dashed lines in Figure \ref{fig:nonmono}) will have two \Minit\ values that correspond 
to the same age and the same EEP. Each point plotted in Figure \ref{fig:nonmono} is taken from one EEP-based track from the dense grid mentioned 
earlier in the paper.  In both cases, even at a resolution of 0.001 \Msun\ the transition is marked by a discontinuity.

One way to avoid this problem is to break the non-monotonic \Minit-age relation into two monotonic relations and then choose one or the
other to represent that EEP. This approach is less than ideal because it requires an arbitrary choice to use one and neglect the other.
Another approach is to artificially smooth the \Minit-age relation before interpolating a new \Minit.

\subsection{Embracing non-monotonic behavior in the \Minit-age relation}\label{ss:double}
The discussion in $\S$\ref{ss:complications} suggests that if we relax the standard mathematical formalism 
for constructing isochrones described above, then we will be able to study phenomena that we otherwise could not. 
The most obvious case is that in which the \Minit-age relation of a given EEP is non-monotonic due to some transition 
in stellar physics over a small range of \Minit.

A simple approach, perhaps the \emph{simplest} approach, is to allow one EEP to represent as many \Minit\ values as the 
\Minit-age relation allows.  In the case of a multi-valued \Minit-age relation we simply divide it into as many monotonic intervals 
as are present and proceed as usual. An isochrone constructed in such a manner will have the \emph{same number of EEPs} but 
a \emph{greater number of $M_{init}$ values} than one constructed in the standard way. Following the example of Figure \ref{fig:nonmono}, a 
multivalued isochrone would have 2 \Minit\ values for the non-monotonic EEP shown.

One challenge with this approach is that the resulting isochrone is itself multi-valued and, therefore, cannot be used in 
the same way as a canonical isochrone. For example, integrating an initial mass function over a canonical isochrone is 
straightforward but the same integral over a multi-valued isochrone is not. This alternative approach to isochrone construction
will require some changes in the way isochrones are used but is nevertheless a worthwhile avenue to explore.


\subsection{Convergence Tests}\label{ss:convergence}

It is informative to consider what level of resolution is needed to faithfully reproduce the predictions of the
underlying grid of stellar evolution tracks in both the transformation from original tracks to EEPs and from EEPs to
isochrones. Here we consider both in turn.

\begin{figure*}
\includegraphics[width=0.6\textwidth]{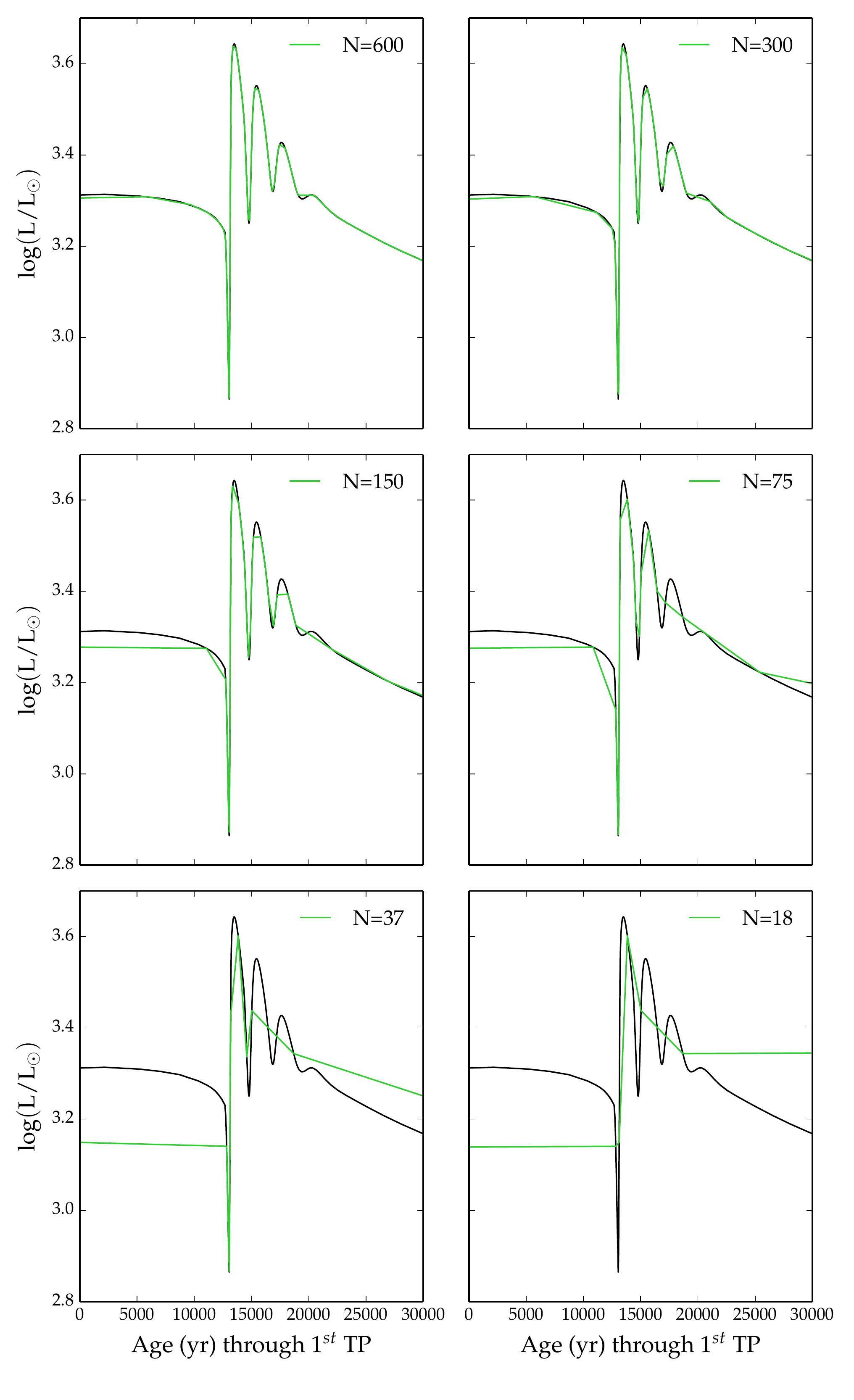}
\caption{Demonstrating the effect of varying the number of secondary EEPs between TP-AGB and PostAGB
on the appearance of the EEP-based track as compared with the original track computed by \MESAstar.
The number of secondary EEPs between the 2 primaries is listed in each panel of the figure.\label{fig:eepTP}}
\end{figure*}

The first is the transformation of stellar evolution tracks to EEP-based tracks as described in $\S$\ref{s:eep}.
The resolution is entirely controlled by the number of secondary EEPs since the number of primary EEP is fixed (and few).
The most important consideration in setting the number of secondary EEPs between any given pair of primary 
EEPs is that the secondaries properly reproduce the morphology of the original tracks. 
In practice, from 50 to 200 secondary EEPs between any pair of primary EEPs should be sufficient to cover smooth phases
of stellar evolution. Here smooth is loosely defined as lacking large derivatives and/or oscillatory behavior of the 
quantities of interest. The worst case is the TP-AGB where the luminosity varies repeatedly by an order of
magnitude or so. To fully resolve the TP-AGB phase in isochrones requires upwards of 500 secondary EEPs between the 
TP-AGB and PostAGB primaries (Conroy et al., {\it Nature}, in press). Consider the example given in Figure \ref{fig:eepTP}
where the luminosity evolution of the original stellar evolution track through one TP is overplotted with a sequence of 
EEP-based tracks for which the number of secondary EEPs ranges from 600 to 18. Without any sort of time-average through
the TP-AGB phase, the eep-based tracks with fewer than 100 EEPs show considerable variations from the original track
and deviate by more than 5\% in the integrated luminosity compared to the original track.

\begin{figure*}
\includegraphics[width=\textwidth]{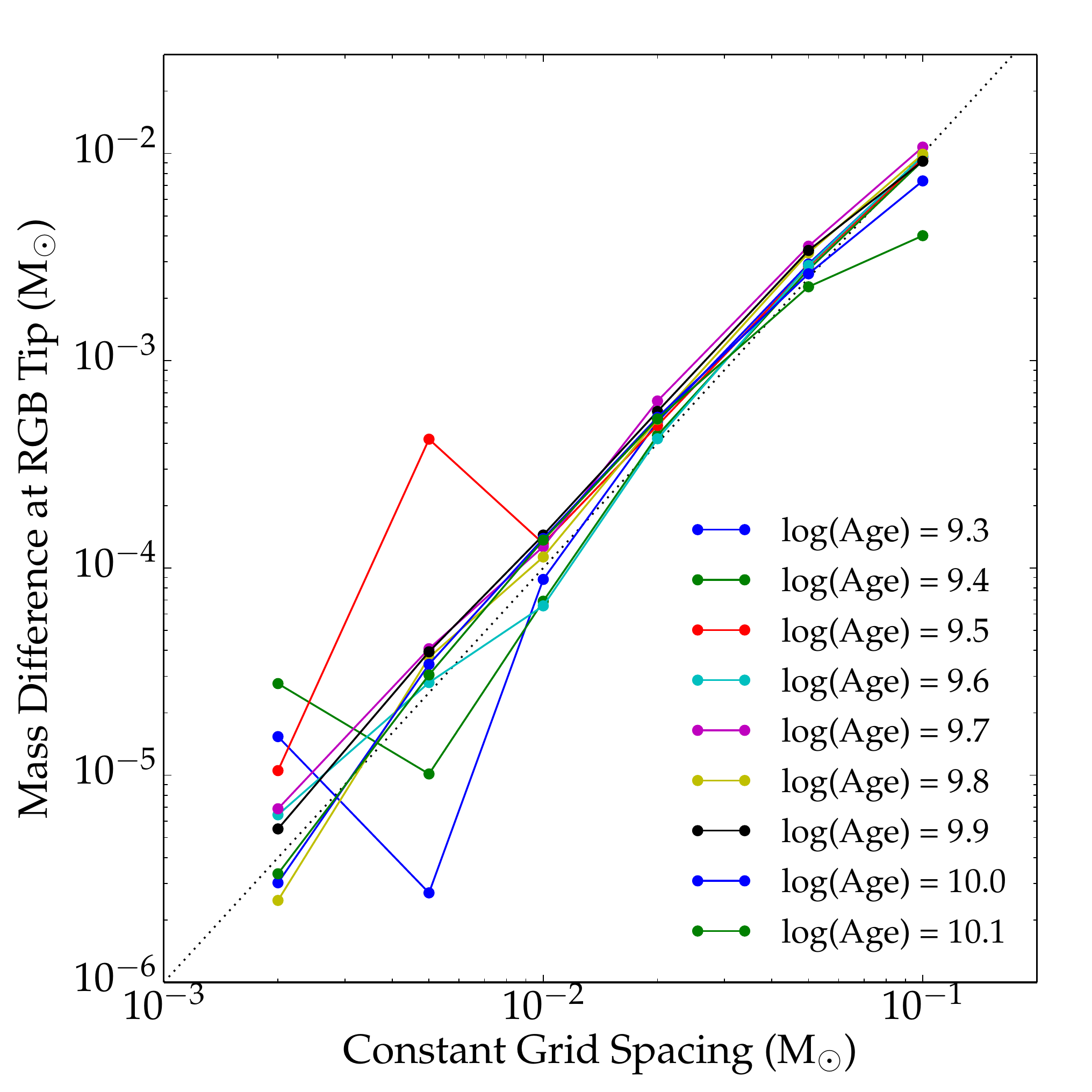}
\caption{The difference in mass at the RGBTip for isochrones constructed from grids with constant mass spacing shown on the x-axis.
The difference is taken with respect to a fiducial grid with mass spacing of 0.001 \Msun\ and the difference is given as an absolute value.
Different ages are shown as different colors; the youngest ages are noisier than the others.
The dotted line shows the square of the mass spacing, which is a decent approximation for the mass difference.\label{fig:iso_conv}}
\end{figure*}

The number of stellar evolution tracks and the \Minit\ spacing is an important consideration in isochrone construction. The following is
a convergence test in which the \Minit\ value reported in isochrones of various grid spacings are compared with a fiducial grid.
The \Minit\ value is the result of the \Minit-age interpolation and, therefore, its value is fundamental in the isochrone construction process.
The fiducial grid of tracks with constant mass-spacing of 0.001 \Msun, mentioned previously, is sampled at successively larger intervals 
of 0.002, 0.005, 0.01, 0.02, 0.05, and 0.1 \Msun. Isochrones with ages between 1 and 10 Gyr are constructed from each of these 
grids and the RGBTip mass is extracted from each isochrone. For a given age, the RGBTip mass from the fiducial grid is compared with the
each of the coarser grids. The absolute mass difference between the fiducial grid and each of the other grids is shown, for a variety of 
ages, in Figure \ref{fig:iso_conv}. The dotted line indicates that the mass difference scales roughly as the square of the \Minit\ spacing.
For example, if we require masses that are reliable at the level of 0.001 \Msun, then a grid with mass spacing of $\sim0.02$ \Msun\ should 
be sufficient.
Whether or not such deviations matter will depend on the application, of course, but it is worthwhile to be able to quantify these effects.

Ultimately, both the total number of EEPs and the \Minit-spacing of the stellar evolution grid should be determined based on the goals
and tolerances to which they will be applied. As a starting place we recommend \Minit-spacing of 0.02 \Msun\ (at least for low-mass stars)
and 1500-2000 EEPs to cover all the phases of evolution considered in $\S$\ref{ss:primary}.

\section{Conclusions}\label{s:conculsions}
A method of transforming a grid of stellar evolution models into isochrones is described in detail. The process is done in two steps.
The first takes the original stellar evolution tracks and converts them onto a uniform basis, suitable for interpolation, called EEPs.
EEPs are divided into two categories: primary, which are physically-defined, and secondary, which divide the interval between two
primary EEPs into a number of equally-spaced points. `Equally-spaced' requires the definition of a metric function that defines a `distance'
along each stellar evolution track. 

The second interpolates within the grid of EEP-based tracks to create both new tracks with different \Minit\ as well as isochrones.
The interpolation requires construction of a relation between \Minit\ and age for each EEP. Generally the \Minit-age relation is monotonic
but in some circumstances (two, in particular) a transition in stellar physics causes the \Minit-age relation to be non-monotonic. 
Non-monotonicity can be dealt with in at least two different ways, which are discussed in the text.

Finally, the extent to which EEP-based tracks and isochrones represent the information in the original grid of stellar evolution tracks is
explored and suggestions for achieving a level of accuracy provided.

\acknowledgments
The author is supported by the Australian Research Council under grant FL110100012.
This paper could not have been written without invaluable contributions from Jieun Choi and Charlie 
Conroy, who provided: many comments on the manuscript, the evolutionary tracks, 
extensive testing of the codes, inspiration, and additional support. Many thanks to Bill Paxton 
and the \MESA\ community for making \MESA\ the wonderful resource that it is.
The isochrone construction process described herein has benefited from discussions with Peter 
Bergbusch, Brian Chaboyer, and Don VandenBerg. Finally, thanks to the anonymous referee for 
providing useful suggestions that improved the clarity of the paper.

\end{document}